\documentclass[aps,prb,twocolumn,superscriptaddress,bibliography]{revtex4-2}

\usepackage{amsfonts}
\usepackage{mathrsfs}
\usepackage{amsmath}
\usepackage{color}
\usepackage{xcolor}
\usepackage{graphicx}
\usepackage{bm}
\usepackage{amssymb}
\usepackage{xspace}
\usepackage{epstopdf}
\usepackage{dcolumn}
\usepackage{longtable}
\usepackage{multirow}
\usepackage{float}
\usepackage{comment}
\usepackage{booktabs}
\usepackage{pifont}

\newcommand{\cmark}{\ding{51}} % ✓
\newcommand{\xmark}{\ding{55}} % ✗
\usepackage[colorlinks=true, letterpaper=true, pdfstartview=FitV,
linkcolor=blue, citecolor=blue, urlcolor=blue]{hyperref}

\makeatother

\begin{document}

\title{Spin and Charge Photocurrents from Electric-Magnetic Dipole Interference in Centrosymmetric Crystals}

\author{Yiwei Zhao}
\affiliation{Department of Physics, City University of Hong Kong, Kowloon 999077, Hong Kong, China}

\author{Jian Zhou}
\thanks{jianzhou@xjtu.edu.cn}
\affiliation{Center for Alloy Innovation and Design, State Key Laboratory for Mechanical Behavior of Materials, Xi'an Jiaotong University, Xi'an 710049, China}

\author{Haowei Xu}
\thanks{haoweixu@cityu.edu.hk}
\affiliation{Department of Physics, City University of Hong Kong, Kowloon 999077, Hong Kong, China}

% \date{\today}

\begin{abstract}
A common view regarding even-order nonlinear optical (NLO) effects, such as second-harmonic generation and the bulk photovoltaic effect, is that the breaking of inversion symmetry is necessary.  However, this view relies on the assumption that light interacts with materials solely through its electric field. Here, we demonstrate that interference between the electric and magnetic components of the same light field can induce even-order NLO effects even in centrosymmetric materials. Using charge and spin photocurrents as prototypical examples, we present a unified microscopic theory showing how such electric-magnetic dipole interference (EMDI) drives NLO responses in inversion-symmetric crystals. Bilinear in both electric and magnetic fields, these responses scale linearly with light intensity. Combining symmetry analysis with first-principles calculations, we decompose the EMDI-induced photocurrents into shift-like and injection-like contributions and classify their dependence on crystal symmetry, time-reversal symmetry, and light polarization.  Notably, we reveal that these EMDI-induced responses in centrosymmetric materials are weaker than conventional responses in non-centrosymmetric materials roughly by a factor of fine structure constant $\alpha \approx [10^{-3}\text{--}10^{-2}] $. Nevertheless, they remain well within experimental reach. Our work thus unlocks NLO functionalities in centrosymmetric materials for e.g., materials characterization, photonics and optoelectronics. This is vital given that many functionally importantly materials, such as cubic Si and cubic $\mathrm{SrTiO_3}$, possess inversion symmetry. 

% implement the theory in first-principles Wannier Hamiltonians and examine centrosymmetric bilayer MoS and magnetic monolayer MnBiTe. The nonmagnetic bilayer provides a clean benchmark of the time-reversal selection rules, whereas MnBiTe demonstrates simultaneous charge and spin photocurrents when time reversal is broken. Momentum-resolved calculations identify the regions of the Brillouin zone responsible for the dominant resonances. 
%Our results establish electric-magnetic dipole interference as a general route to coupled charge and spin photocurrents in centrosymmetric quantum materials and provide a symmetry framework for separating relativistic and magnetic contributions.
\end{abstract}

\maketitle

\emph{Introduction.} Even-order nonlinear optical (NLO) effects are central for many technologies, such as photodetection, energy conversion, and the optical characterization of quantum materials~\cite{alex2011tip,bhatnagar2013role,spanier2016power,osterhoudt2019colossal,ma2019nonlinear}. A prototypical example of second-order NLO effects is the bulk photovoltaic effect (BPVE)~\cite{fridkin2001bulk,glass1995high,von1981theory,sipe1993nonlinear,bhalla2023quantum}, by which dc currents can be generated under optical excitation. Its two canonical microscopic mechanisms, shift and injection currents, encode interband coherence, band velocities, and quantum geometry ~\cite{sipe2000second,young2012first,nastos2010optical,dejuan2017quantized,ahn2020low,ahn2022riemannian,ma2021topology,ma2023photocurrent}. The prevailing view is that these even-order NLO effects strictly require inversion symmetry breaking~\cite{zhang2019symmetry,cook2017design,morimoto2016topological,tan2016shift,taghinejad2020transient,morimoto2018current,young2012first}, restricting them to non-centrosymmetric materials. This requirement severely limits the applicability of even-order NLO effects, given that many technologically crucial materials, such as silicon, are centrosymmetric.

To some extent, however, this restriction is a misconception. It arises because traditionally only the electric field $E$ of the light is considered for light-matter interaction. As a polar vector, $E$ reverses sign under spatial inversion, which forbids many even-order NLO responses, including conventional second-harmonic generation and bulk photovoltaic effect, in centrosymmetric materials. Nevertheless, light also possesses a magnetic field $B$, which behaves as an axial vector and remains invariant under spatial inversion. Incorporating the magnetic field would fundamentally alter the symmetry restrictions governing NLO responses.

In this work, we demonstrate that the interference between the electric and magnetic components of the same light field can trigger even-order NLO effects that are conventionally forbidden by inversion symmetry. Specifically, we show that such electric-magnetic dipole interference (EMDI) can generate charge and spin photocurrents ($j \propto BE$) in centrosymmetric materials. These EMDI-induced photocurrents can be separated into shift-like and injection-like contributions across four distinct response channels~\cite{xu2021pure}: $L$-shift, $C$-shift, $L$-injection, and $C$-injection, where $L$ and $C$ denote the in-phase and quadrature sectors of the light field. While all four channels are even under spatial inversion ($\mathcal{P}$), time-reversal symmetry ($\mathcal{T}$) selectively classifies them into even and odd sectors that are complementary for charge and spin photocurrents. Using both order-of-magnitude estimations and first-principles calculations, we show that EMDI-induced photocurrents are weaker than the conventional bulk charge and spin photocurrents in noncentrosymmetric materials by a factor of $\alpha \approx [10^{-3}\text{--}10^{-2}] $, where $\alpha$ is the fine structure constant. Nevertheless, they remain readily accessible in experiments.  Our results thus establish a unified theory for light-driven charge and spin currents in centrosymmetric materials. These findings expand the reach of diverse NLO technologies, such as materials characterization, photonics, optoelectronics, to centrosymmetric materials.

\emph{Microscopic theory of EMDI-induced photocurrents.} We consider a monochromatic light field with angular frequency $\omega$. The light-matter interaction contains both electric-dipole and spin magnetic-dipole (Zeeman) components, that is,
\begin{equation}{\label{eq:light-matter}}
\begin{aligned}
\hat V(t) & = \hat V_E(t) + \hat V_B(t) \\
& =
-e\,\mathbf E(t)\cdot\hat{\mathbf r}
-\mu_B\,
\mathbf B(t)\cdot\hat{\bm\sigma},
\end{aligned}
\end{equation}
where $e$ and $\mu_B$  are the electron charge and the Bohr magneton, while $\hat{\bm\sigma}$ are Pauli matrices acting in spin space. While previous studies focused almost exclusively on $\hat V_E$, here we harness the interference between $\hat V_E$ and $\hat V_B$ to drive unconventional NLO responses. Although EMDI can induce diverse NLO responses, we focus specifically on photocurrents., which can be expressed as 
\begin{equation}
J^{a, s_d}
=
\chi_{abc}^{s_d}E_bB_c,
\label{eq:response}
\end{equation}
where $a, b, c$ are Cartesian indices. $s_d$ denotes the spin polarization for spin photocurrents, and is omitted for  charge photocurrent.

Generally speaking, $J$ results from the second-order perturbation $\Delta \rho^{(2)}$ of the electron density matrix under light-matter interaction Eq.~(\ref{eq:light-matter}). For EMDI-induced responses, $\Delta \rho^{(2)}$ is first-order in both electric and magnetic fields, making the overall response second-order in the optical fields.  Detailed derivations for $\Delta \rho^{(2)}$ can be found in the Supplemental Material (SM). It is instructive  to decompose the EMDI-induced photocurrents into an in-phase sector ($L$), and a quadrature sector ($C$). Besides, photocurrent arise from two principal mechanisms: shift-like and injection-like. Consequently, the total photocurrent comprises four distinct channels. For charge photocurrents, the conductivities are given by:
\begin{equation}\label{eq:conductivities}
    \begin{aligned}
\chi_{abc}^{L,\mathrm{sh}}
&=
\mathcal C
\int[d\mathbf k]\,
f_{nm}\,
\Im\!\left[
\sigma_{mn}^{c}r_{nm;a}^{b}
+
r_{mn}^{b}\sigma_{nm;a}^{c}
\right]
\delta(\omega-\omega_{nm}),
% \label{eq:Lshift}
\\
\chi_{abc}^{C,\mathrm{sh}}
&=
\mathcal C
\int[d\mathbf k]\,
f_{nm}\,
\Re\!\left[
\sigma_{mn}^{c}r_{nm;a}^{b}
-
r_{mn}^{b}\sigma_{nm;a}^{c}
\right]
\delta(\omega-\omega_{nm}),
% \label{eq:Cshift}
\\
\chi_{abc}^{L,\mathrm{inj}}
&=
\mathcal C
\int[d\mathbf k]\,
f_{nm}\,
\Delta v_{nm}^{a}\,
\Re\!\left[
r_{nm}^{b}\sigma_{mn}^{c}
\right]
\delta(\omega-\omega_{nm}),
% \label{eq:Linjection}
\\
\chi_{abc}^{C,\mathrm{inj}}
&=
-\mathcal C
\int[d\mathbf k]\,
f_{nm}\,
\Delta v_{nm}^{a}\,
\Im\!\left[
r_{nm}^{b}\sigma_{mn}^{c}
\right]
\delta(\omega-\omega_{nm}).
% \label{eq:Cinjection}
\end{aligned}
\end{equation}
The conductivities for spin photocurrents are given in the SM. Here, $\mathcal C = \frac{\pi e^2 g\mu_B}{\hbar^2}$ is a pre-factor. $m,n$ and $\mathbf{k}$ are electron band indices and wavevector. , $r_{mn}$, $v_{mn}$, and $\sigma_{mn}$ are position, velocity, and spin matrix elements, respectively. $r_{nm;a}^{b}$ and $\sigma_{nm;a}^{c}$ are the generalized derivatives, which ensures gauge covariance (SM). Meanwhile, $f_{nm}=f_n-f_m$ and $\Delta v_{nm}^{a}=v_{nn}^{a}-v_{mm}^{a}$ are the difference in electron occupation numbers and band velocities. Summation over repeated band indices is implied. The superscripts $\rm sh$ and $\rm inj$ denotes shift-like and injection-like responses, respectively. Note that the injection-like conductivity represents a current-injection rate, and the stationary injection current should contain an additional phenomenological relaxation time $\tau$ ~\cite{sipe2000second,matsyshyn2021rabi}.

% The Its off-diagonal component gives rise to the shift-like photocurrent, whereas the diagonal population correction produces the injection-like response.

%The dc response originates from the mixed second-order density matrix containing one electric-dipole and one spin magnetic-dipole interaction. 
% We evaluate the mixed electric-magnetic response within the length-gauge density-matrix formalism \cite{sipe2000second}. 

% For $n\neq m$, the interband position matrix element is
% \begin{equation}
% r_{nm}^{b}
% =
% (1-\delta_{nm})\xi_{nm}^{b}
% =
% \frac{v_{nm}^{b}}{i\omega_{nm}},
% \label{eq:rnm}
% \end{equation}
% where
% $\omega_{nm}=(\varepsilon_n-\varepsilon_m)/\hbar$.

For a single incident light beam, the total EMDI photocurrent is obtained by combining the shift-like and injection-like contributions associated with the corresponding electric- and magnetic-field components. For linearly polarized light propagating along $+z$, we choose
\begin{equation}
\mathbf{E}=(E_0,0,0),\qquad
\mathbf{B}=(0,B_0,0),
\end{equation}
with $B_0=E_0/c$. The resulting photocurrent is
\begin{equation}
J^a
=
\left(
\chi_{axy}^{L,\mathrm{sh}}
+
\tau\chi_{axy}^{L,\mathrm{inj}}
\right)E_0B_0 .
\label{eq:total_LP}
\end{equation}
For circularly polarized light propagating along the same direction, we write
\begin{equation}
\mathbf{E}
=
\frac{E_0}{\sqrt{2}}(1,is,0),
\qquad
\mathbf{B}
=
\frac{B_0}{\sqrt{2}}(-is,1,0),
\end{equation}
where $s=\pm1$ denotes the light helicity. The total photocurrent is then
\begin{equation}
\begin{aligned}
J^{a}
=
\frac{E_{0}B_{0}}{2}
\Big[
&
\left(
\chi_{axy}^{\mathrm{L,sh}}
+
\tau\chi_{axy}^{\mathrm{L,inj}}
\right)
+
\left(
\chi_{ayx}^{\mathrm{L,sh}}
+
\tau\chi_{ayx}^{\mathrm{L,inj}}
\right)
\\
&
+s\left(
\chi_{axx}^{\mathrm{C,sh}}
+
\tau\chi_{axx}^{\mathrm{C,inj}}
\right)
-s\left(
\chi_{ayy}^{\mathrm{C,sh}}
+
\tau\chi_{ayy}^{\mathrm{C,inj}}
\right)
\Big].
\end{aligned}
\label{eq:total_CP}
\end{equation}

These expressions also clarify the physical distinction between the $L$ and $C$ sectors. For linearly polarized light (LPL), the electric and magnetic fields are in phase and mutually perpendicular, so only the $L$ sector contributes and the relevant conductivity components satisfy $b\perp c$. For circularly polarized light (CPL), the complex electric- and magnetic-field components contain relative quadrature phases, giving rise to the helicity-dependent $C$ sector in addition to the $L$ sector. Consequently, although the instantaneous electric and magnetic fields remain perpendicular for a plane wave, the products of their complex Cartesian components can generate finite contributions with $b\parallel c$, as explicitly seen in Eq.~(\ref{eq:total_CP}). Reversing the light helicity, $s\rightarrow -s$, changes the sign of the $C$-sector contribution while leaving the $L$-sector contribution unchanged.

For a spin photocurrent flowing along $a$ with spin polarization along $d$, we employ the standard  spin-current operator~\cite{xu2021pure}
$
\hat J^{a_d}
=
\frac{1}{2}
\left\{
\hat v^{a},
\hat s_d
\right\}.
$ Like charge photocurrents, the spin photocurrents can also be decomposed into shift-like and injection-like mechanisms across $L$ and $C$ sectors, yielding four distinct channels analogous  to
Eqs.~(\ref{eq:conductivities}) (details in SM). However, owing to the spin degree of freedom, the symmetry selection rules for spin photoresponses differ fundamentally from those governing charge responses \cite{xu2021pure,lihm2022comprehensive}.

A crucial aspect of EMDI-induced responses is their magnitude relative to conventional photoresponses in non-centrosymmetric  materials. Because the electric-dipole interaction $erE$ is complemented by the magnetic dipole term $\mu_B B$, the EMDI response is weaker than standard non-centrosymmetric processes by a factor of 
\begin{equation}\label{eq:magnitude_comparison}
    \mathcal{A} \sim \frac{\mu_B} {ea_0 c_0} \sim \alpha \approx [10^{-3}\text{--}10^{-2}]
\end{equation}
where $a_0$ is thr Bohr radius, $\alpha$ is the fine-structure constant, and $E=c_0 B$ relates the electric and magnetic components of a light field with $c_0$ the speed of light. Here, we used the fact that the electric dipole moment of band electrons is on the order of $ea_0$ 

\begin{table}[]
\centering
\caption{
Transformation properties of the light-induced mixed electric--magnetic
charge and spin photocurrent tensors under spatial inversion
$\mathcal{P}$, time reversal $\mathcal{T}$, and the combined operation
$\mathcal{PT}$.
Here $d$ denotes the spin-polarization direction for spin photocurrents.
}
\label{tab:symmetry_ME_photocurrent}

\footnotesize
\renewcommand{\arraystretch}{1.35}
\setlength{\tabcolsep}{6pt}

\begin{tabular*}{\columnwidth}{
@{\extracolsep{\fill}}
c
c
c
c
}

\specialrule{0.10em}{0pt}{0.10em}
\specialrule{0.04em}{0pt}{0.25em}

Response
&
$\mathcal{P}$
&
$\mathcal{T}$
&
$\mathcal{PT}$
\\

\midrule

$\chi_{abc}^{\mathrm{L,sh}}$
&
\cmark
&
\xmark
&
\xmark
\\

$\chi_{abc}^{\mathrm{C,sh}}$
&
\cmark
&
\cmark
&
\cmark
\\

$\chi_{abc}^{\mathrm{L,inj}}$
&
\cmark
&
\cmark
&
\cmark
\\

$\chi_{abc}^{\mathrm{C,inj}}$
&
\cmark
&
\xmark
&
\xmark
\\

\midrule

$\chi_{abc}^{\mathrm{L,sh},s_d}$
&
\cmark
&
\cmark
&
\cmark
\\

$\chi_{abc}^{\mathrm{C,sh},s_d}$
&
\cmark
&
\xmark
&
\xmark
\\

$\chi_{abc}^{\mathrm{L,inj},s_d}$
&
\cmark
&
\xmark
&
\xmark
\\

$\chi_{abc}^{\mathrm{C,inj},s_d}$
&
\cmark
&
\cmark
&
\cmark
\\

\specialrule{0.04em}{0.25em}{0pt}
\specialrule{0.10em}{0.10em}{0pt}

\end{tabular*}
\end{table}

\emph{Symmetry analyses.} Symmetries are crucial for NLO responses. Table~\ref{tab:symmetry_ME_photocurrent} summarizes the symmetry properties of the EMDI-induced charge and spin photocurrent tensors. In contrast to the conventional electric-dipole second-order photovoltaic response, the EMDI-induced spin and charge photocurrents are even under spatial inversion $\mathcal P$. All four channels can  remain finite in centrosymmetric crystals. 

Interestingly, time reversal $\mathcal T$ further separates the four charge-current channels into two complementary sectors. The $L$-shift and $C$-injection responses are $\mathcal T$-odd, whereas the $C$-shift and $L$-injection responses are $\mathcal T$-even. Consequently, only the latter two channels  survive for EMDI-induced charge photocurrents when time-reversal symmetry is preserved. For spin photocurrents, the classification is exactly reversed. This is because in contrast to charge current, the spin current $\hat J^{a_d}=\frac{1}{2}\{\hat v^a,\hat s_d\}$ is $\mathcal T$ even. Consequently, it can be shown that the $L$-shift and $C$-injection channels are  $\mathcal T$-even for spin photocurrents, whereas the $C$-shift and $L$-injection channels are $\mathcal T$-odd. This constitutes a central distinction for EMDI-induced charge and spin photocurrents.

In addition, the EMDI responses are governed by other crystal symmetries as well. Here we take mirror symmetry as an example. For a mirror plane normal to the $i$-th direction, the charge responses transform with the factor $(-1)^{\delta_{ia}+\delta_{ib}+\delta_{ic}+1}$, whereas the spin responses acquire the additional spin-index contribution and transform as $(-1)^{\delta_{ia}+\delta_{ib}+\delta_{ic}+\delta_{id}}$. We can also consider the influence of other combined symmetries like $\mathcal{P} \mathcal{T}$ and $\mathcal M_i\mathcal T$, and the results are summarized  in Table~\ref{tab:symmetry_ME_photocurrent}. 

\begin{figure}[]
\begin{centering}
\includegraphics[width=8.4cm]{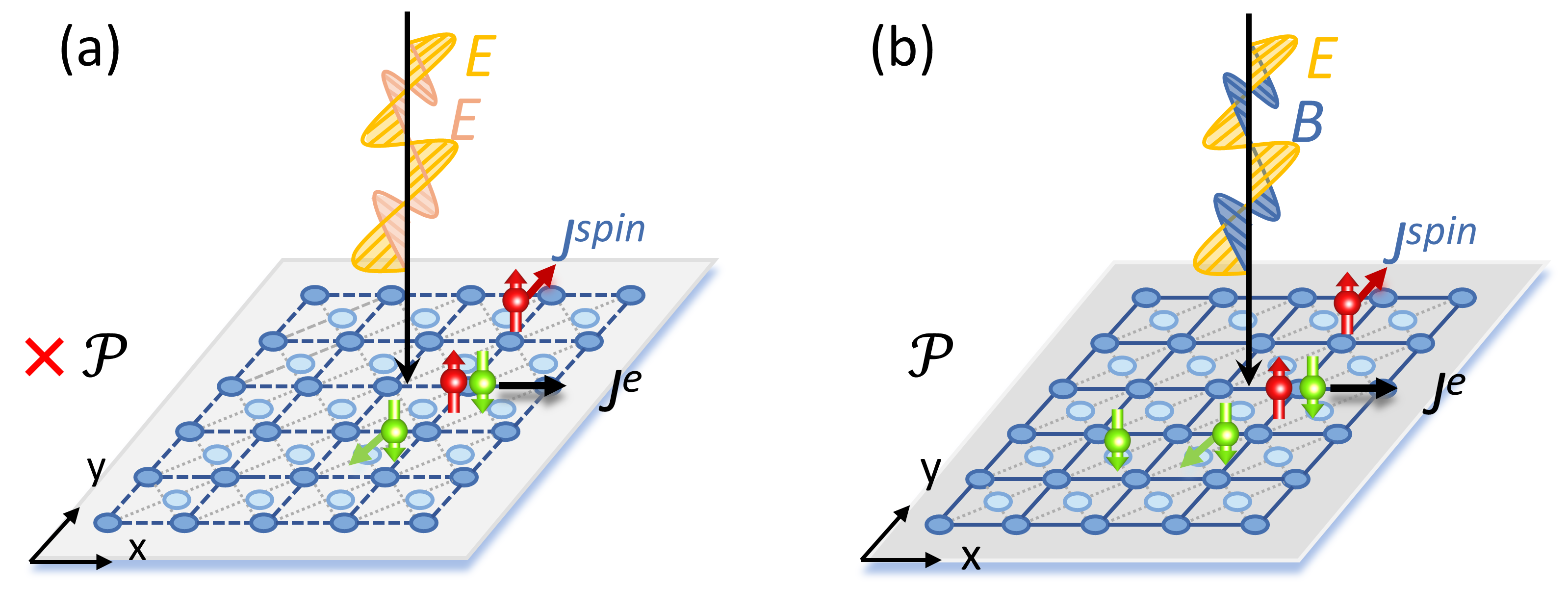}
\par\end{centering}
\caption{\label{Fig_1}
Schematic illustration of conventional electric-dipole and mixed electric-magnetic photogalvanic responses.
(a) In a noncentrosymmetric system with broken inversion symmetry $\mathcal P$ but preserved mirror symmetry $\mathcal M_y$, the conventional electric-dipole channel can generate charge current $J^e$ and spin current $J^{\rm spin}$. The right panel illustrates a representative bilayer lattice with two inequivalent sites (A and B).
(b) Restoring inversion symmetry suppresses the conventional electric-dipole second-order dc response.
(c) The magnetic component $\mathbf B$ of the same optical wave opens an electric-magnetic interference channel. Because $\mathbf E$ is polar whereas $\mathbf B$ and spin are axial, finite charge and spin photocurrents can be generated without breaking inversion symmetry. Black arrows denote charge currents and colored arrows denote spin-polarized currents.
}
\end{figure}

\emph{Centrosymmetric bilayer MoS$_2$} Next, we illustrate the EMDI-induced photocurrents in some realistic materials. We first consider centrosymmetric bilayer MoS$_2$, whose magnetic point group is $\bar{3}m1'$. This grey magnetic point group contains both spatial inversion $\mathcal P$ and time reversal $\mathcal T$ as independent symmetries, together with the threefold rotation $C_{3z}$ and vertical mirror operations. 

% The allowed response channels can first be identified directly from their time-reversal properties summarized in Table~\ref{tab:symmetry_ME_photocurrent}. For the shift-like response,
% \begin{equation}
% \chi_{abc}^{L,\mathrm{sh}}
% \xrightarrow{\mathcal T}
% -\chi_{abc}^{L,\mathrm{sh}},
% \qquad
% \chi_{abc}^{C,\mathrm{sh}}
% \xrightarrow{\mathcal T}
% +\chi_{abc}^{C,\mathrm{sh}},
% \end{equation}
% whereas for the injection-like response,
% \begin{equation}
% \chi_{abc}^{L,\mathrm{inj}}
% \xrightarrow{\mathcal T}
% +\chi_{abc}^{L,\mathrm{inj}},
% \qquad
% \chi_{abc}^{C,\mathrm{inj}}
% \xrightarrow{\mathcal T}
% -\chi_{abc}^{C,\mathrm{inj}}.
% \end{equation}
Because $\mathcal T$ is preserved in MoS$_2$, all time-reversal-odd charge responses must vanish identically, that is,
\begin{equation}
\chi_{abc}^{L,\mathrm{sh}}
=
\chi_{abc}^{C,\mathrm{inj}}
=
0
\end{equation}
leaving only the $C$-shift and $L$-injection channels for charge photocurrents. For spin photocurrents, however, the $C$-shift and $L$-injection channels survive while the $L$-shift and $C$-injection channels are forbidden. Other crystal symmetries further constrain non-zero components of the photo-conductivities, which is analyzed in the SM.

\begin{figure*}[]
\begin{centering}
\includegraphics[width=0.75\textwidth]{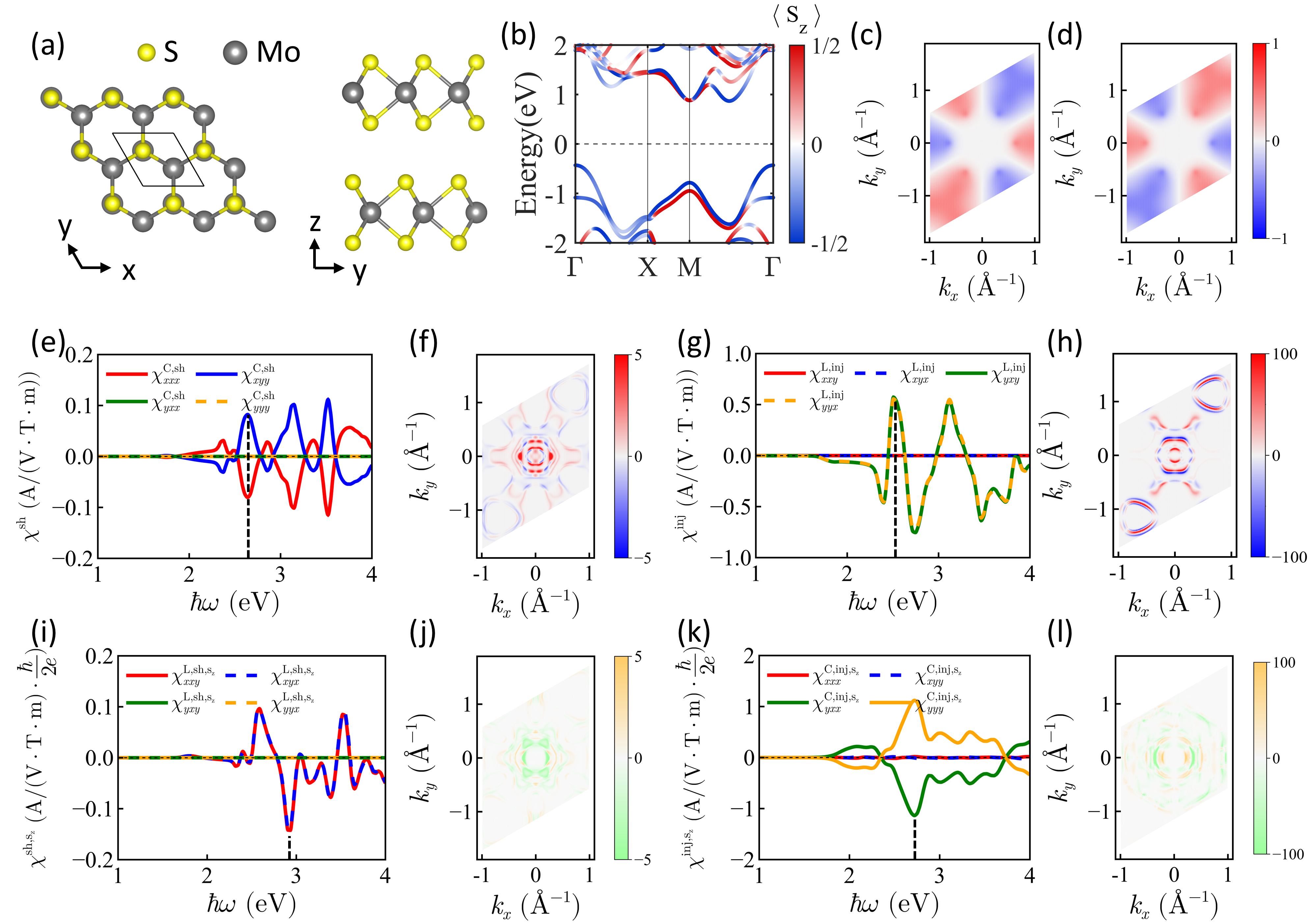}
\par\end{centering}
\caption{\label{Fig_2}
Mixed electric--magnetic dipole photogalvanic responses in centrosymmetric bilayer MoS$_2$.
(a) Top and side views of the centrosymmetric 2H-stacked bilayer MoS$_2$ structure.
(b) Band dispersion along the high-symmetry $\mathbf{k}$ path, where the color scale represents the spin polarization $\langle S_z\rangle$. Spin--orbit coupling (SOC) is included in the calculation. (c,d) The spin-$z$ texture $s^{z}_{mm}(\mathbf{k})$ of the highest valence band of MoS$_2$ for (c) the top layer and (d) the bottom layer, respectively. (e) Circular/quadrature shift-current ($C$-sh) spectra for representative tensor components. (f) Momentum-resolved distribution of the $yxy$ component, $\chi_{yxy}^{C,\mathrm{sh}}(\mathbf{k},\omega)$, at the photon energy marked by the vertical dashed line in (e). (g) Linear/in-phase injection-current ($L$-inj) spectra for representative tensor components. (h) Momentum-resolved distribution of the $yxy$ component, $\chi_{yxy}^{L,\mathrm{inj}}(\mathbf{k},\omega)$, at the photon energy marked by the vertical dashed line in (g).
(i) Linear/in-phase spin shift-current ($L$-sh) spectra for representative spin-current tensor components, where the spin-current response is expressed in units of $\hbar/2e$ relative to the corresponding charge-current units. (j) Momentum-resolved distribution of the $yxy$ component of the $L$-sh spin-current response, $\chi_{yxy}^{L,\mathrm{sh}}(\mathbf{k},\omega)$, at the photon energy marked by the vertical dashed line in (i).
(k) Circular/quadrature spin injection-current ($C$-inj) spectra for representative spin-current tensor components.
(l) Momentum-resolved distribution of the $yxy$ component of the $C$-inj spin-current response, $\chi_{yxy}^{C,\mathrm{inj}}(\mathbf{k},\omega)$, at the photon energy marked by the vertical dashed line in (k).
}
\end{figure*}

\emph{Centrosymmetric bilayer MoS$_2$.}---We now demonstrate the EMDI photocurrents in a realistic centrosymmetric material. We first consider 2H-stacked bilayer MoS$_2$ [Fig.~\ref{Fig_2}(a)], whose magnetic point group is $\bar{3}m1'$. This grey magnetic point group contains both spatial inversion $\mathcal P$ and time reversal $\mathcal T$ as independent symmetries. Figure~\ref{Fig_2}(b) shows the band dispersion including spin--orbit coupling, with the color scale denoting the out-of-plane spin polarization $\langle S_z\rangle$. Bilayer MoS$_2$ is an indirect-gap semiconductor with a band gap of approximately $1.31$~eV, with the valence-band maximum located at $\Gamma$ and the conduction-band minimum at the $M$ point. Although the bilayer is globally inversion symmetric, the two inversion-related layers carry opposite local spin textures. This hidden spin polarization is directly seen in the layer-resolved $s_{mm}^{z}(\bm{k})$ distributions of the highest valence states shown in Figs.~\ref{Fig_2}(c) and \ref{Fig_2}(d), where the spin textures of the top and bottom layers are reversed with respect to each other.

The presence of $\mathcal T$ sharply separates the four EMDI channels introduced above. For charge photocurrents, the $L$-shift and $C$-injection responses are odd under $\mathcal T$ and therefore vanish identically, $\chi_{abc}^{L,\mathrm{sh}}=\chi_{abc}^{C,\mathrm{inj}}=0$, whereas the $C$-shift and $L$-injection channels are allowed. The calculated $C$-shift spectra are shown in Fig.~\ref{Fig_2}(e). Representative in-plane tensor components clearly separate into symmetry-allowed and forbidden sets: $\chi_{xxx}^{C,\mathrm{sh}}$=- $\chi_{xyy}^{C,\mathrm{sh}}$, whereas $\chi_{yxx}^{C,\mathrm{sh}}$ and $\chi_{yyy}^{C,\mathrm{sh}}$ vanishes. Figure~\ref{Fig_2}(f) further resolves the $\chi_{yxy}^{C,\mathrm{sh}}$  component in momentum space at $\hbar\omega=2.64$~eV, which suggests that the current is mainly contributed from the vicinity of the $\Gamma$ point. The $L$-injection response, is shown in Fig.~\ref{Fig_2}(g). The momentum-resolved distribution of $\chi_{yxy}^{L,\mathrm{inj}}(\bm{k},\omega)$ at $\hbar\omega=2.50$~eV is displayed in Fig.~\ref{Fig_2}(h). At $\hbar\omega=3.12$~eV, the calculated charge responses are
$\tau\chi_{yxy}^{\mathrm{L,inj}}
=\tau\chi_{yyx}^{\mathrm{L,inj}}
=0.62~\mathrm{A/(V\,T\,m)}$
and
$\chi_{xyy}^{\mathrm{C,sh}}
=-\chi_{xxx}^{\mathrm{C,sh}}
=0.12~\mathrm{A/(V\,T\,m)}$.
Hence, according to  Eqs.~(\ref{eq:total_LP}) an $x$-polarized LPL at this photon energy with an alternating electric-field magnitude of $E=0.5~\mathrm{V/nm}$ could trigger a current density of $5.18\times10^{-4}~\mu\mathrm{A/nm^2}$ along $+y$. Under CPL with the same field strength, the $L$-injection response produces a helicity-independent $J_y\simeq5.18\times10^{-4}~\mu\mathrm{A/nm^2}$, whereas the $C$-shift response gives a helicity-dependent transverse current $J_x\simeq-1.00\times10^{-4}~\mu\mathrm{A/nm^2}$.

For spin photocurrents, the $L$-shift and $C$-injection spin photocurrents are allowed, whereas the $C$-shift and $L$-injection spin channels are forbidden. Figure~\ref{Fig_2}(i) shows the calculated $L$-shift spin-photocurrent spectra. The momentum-resolved distribution of the $xxy$ component at $\hbar\omega=2.91$~eV is shown in Fig.~\ref{Fig_2}(j), revealing the spin current is mainly contributed from the vicinity of the $\Gamma$ point. Figs.~\ref{Fig_2}(k) and \ref{Fig_2}(l) show the corresponding $C$-injection spin photocurrent and its momentum-space distribution at $\hbar\omega=2.72$~eV, respectively. Remarkably, at $\hbar\omega=2.72$~eV, the calculated 
$\chi_{yyy}^{\mathrm{C,inj,s_{z}}}
=-\chi_{yxx}^{\mathrm{C,inj,s_{z}}}
=1.00~\mathrm{A/(V\,T\,m)}$.
Hence, under CPL with the same field strength, the spin photocurrent is approximately one order of magnitude larger than the charge photocurrent.
Together, these results demonstrate that inversion symmetry does not suppress the mixed-$EB$ nonlinear response, but instead, together with time reversal and the remaining crystal symmetries, selects distinct charge- and spin-photocurrent channels.

\begin{figure*}[]
\begin{centering}
\includegraphics[width=0.75\textwidth]{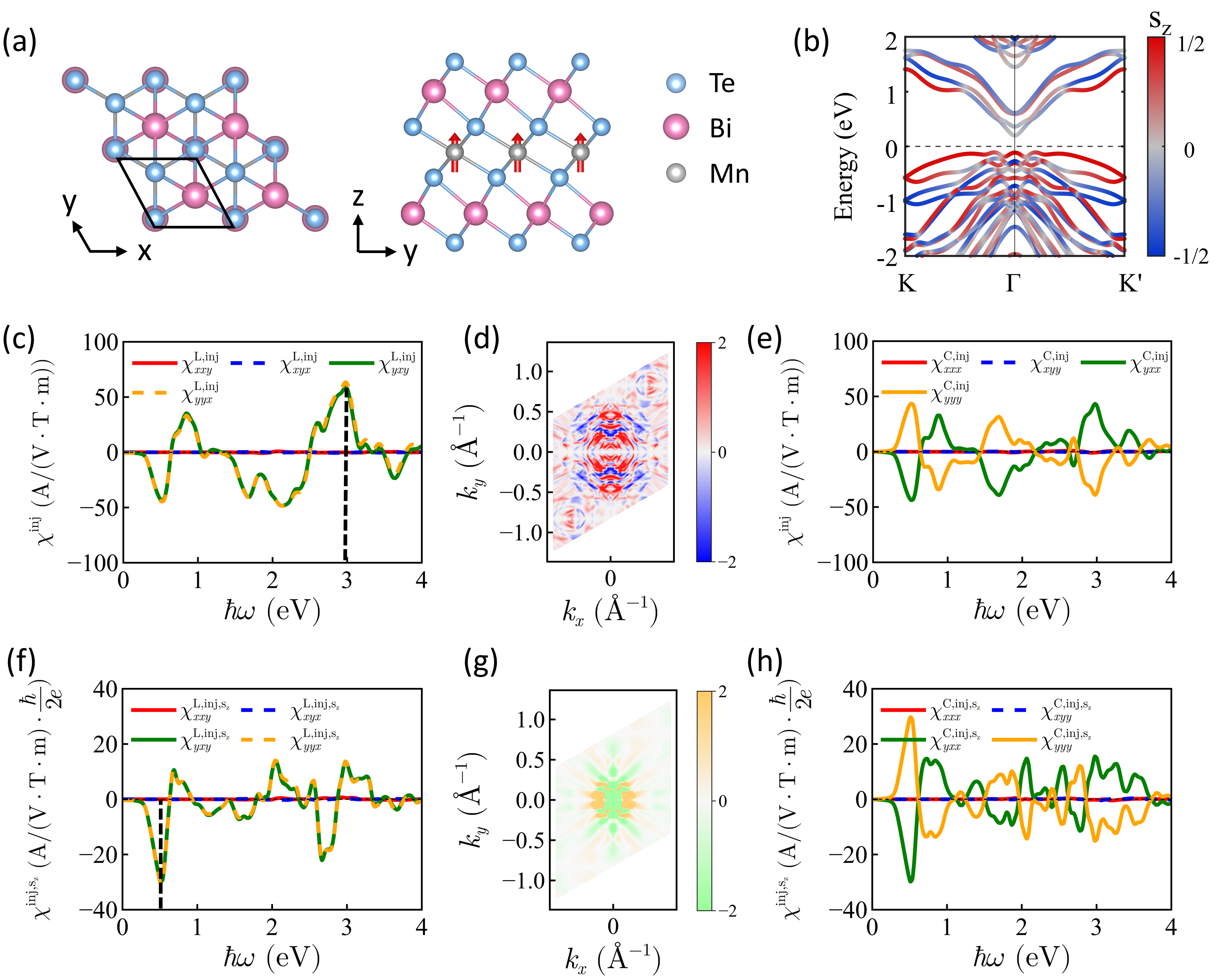}
\par\end{centering}
\caption{\label{Fig_3}
Charge and spin photocurrents generated by EMDI in monolayer MnBi$_2$Te$_4$.
(a) Top and side views of the crystal structure; red arrows denote Mn magnetic moments.
(b) Spin-resolved electronic band structure along the $K$-$M$-$K'$ path, with color indicating $\langle S_z\rangle$.
(c) Calculated  $L$- and (e) $C$-inject charge current response for representative tensor components.
(d) Momentum-resolved charge $L$-inject response at the photon energy marked by the vertical dashed line in (c).
(f) Calculated $L$- and (h) $C$-inject spin current response.
(g) Corresponding momentum-resolved spin $L$-inject response at the energy marked in (f).
}
\end{figure*}

\emph{Centrosymmetric magnetic MnBi$_2$Te$_4$} We next turn to monolayer MnBi$_2$Te$_4$, a centrosymmetric magnetic system in which $\mathcal{T}$ is broken while $\mathcal{P}$ remains intact. Figure~\ref{Fig_3}(a) shows the top and side views of the MnBi$_2$Te$_4$ layer. Each Mn atom carries a magnetic moment of approximately $5\,\mu_{\mathrm{B}}$, pointing along the easy-axis $z$ direction. Figure~\ref{Fig_3}(b) shows the calculated electronic band structure, exhibiting semiconducting behavior with a direct band gap of $0.31$~eV. 

We next analyze the photoconductivities under symmetry constraints. A key consequence of broken $\mathcal{T}$ is that the $\mathcal{T}$-selection rules, which forbid half of the channels as in the case of  bilayer MoS$_2$, are lifted. Consequently, all four channels can appear regardless of their parity under $\mathcal{T}$-operation. With the Mn moments aligned along the out-of-plane $z$ direction, monolayer MnBi$_2$Te$_4$ belongs to the magnetic point group $\bar{3}m'$. The threefold rotation $C_{3z}$ and spatial inversion $\mathcal P$ remain unitary symmetries, whereas the vertical mirror $\mathcal{M}_x$ in the atomic structure reverses the axial magnetic moment $M_z$ and is therefore broken. Nevertheless, since time reversal $\mathcal T$ also reverses $M_z$, the combined antiunitary symmetry $\mathcal{M}_x\mathcal T$ is preserved.

Interestingly, for $\mathcal T$-even sectors ($L$-injection and $C$-shift channels) for charge photocurrents, the non-vanishing components of the photocurrent conductivity tensors are $
\chi_{xxx}
=
-\chi_{xyy}
=
-\chi_{yxy}
=
-\chi_{yyx}
\label{eq:MBT_even_components}
$, while all other components vanish. Conversely, for the $\mathcal T$-odd sector ($L$-shift and $C$-injection channels) , it is the $
\chi_{xxy}
=
\chi_{xyx}
=
\chi_{yxx}
=
-\chi_{yyy}
\label{eq:MBT_odd_components}$ components that are nonzero.  Figures~\ref{Fig_3}(c) and \ref{Fig_3}(e) show the $L$- and $C$-injection responses, respectively. The symmetry-dictated pattern can be clearly observed. The corresponding momentum-resolved distributions in Figs.~\ref{Fig_3}(d)  further reveal how the symmetry-allowed response is built from positive and negative contributions over the Brillouin zone. Meanwhile, Figures~\ref{Fig_3}(f) and \ref{Fig_3}(h) show the calculated $L$- and $C$-injection spin photocurrents. Their magnitudes are comparable to the corresponding charge responses after expressing the spin current in the conventional $\hbar/2e$ units. The $k$-space distribution in Fig.~\ref{Fig_3}(g) further shows that the spin response is predominantly concentrated around the $\Gamma$ point.

For completeness, the calculated shift-current responses are provided in the Supplemental Material, while here we focus on the substantially larger injection-current contributions. Notably, the $y$-directed photocurrent generated by an $x$-polarized LPL is entirely contributed by the $L$-injection channel. At $\hbar\omega=3.0$~eV, the calculated response reaches $\tau\chi_{yxy}^{\mathrm{L,inj}}\simeq60~\mathrm{A/(V\,T\,m)}$. Substituting this value into Eq.~(\ref{eq:total_LP}), an incident LPL with an electric-field amplitude of $E_0=0.5~\mathrm{V/nm}$, with $B_0=E_0/c$, yields a charge-current density of approximately $5.0\times10^{-2}~\mu\mathrm{A/nm^2}$ along the $y$ direction. The corresponding spin-injection response reaches approximately $30~\mathrm{A/(V\,T\,m)}$ in units of $\hbar/2e$, giving a spin-current density of about $2.5\times10^{-2}~\mu\mathrm{A/nm^2}$. Remarkably, these current densities are only about two orders of magnitude smaller than representative bulk-photovoltaic current densities reported in two-dimensional materials, demonstrating that the EMDI-induced photocurrent can reach an experimentally relevant magnitude despite originating from the much weaker magnetic-dipole coupling.

\emph{Discussions and Conclusion.} Our EMDI photocurrents should be distinguished from the previously studied magnetic BPVE. In magnetic proximity structures, magnetic order can reduce the effective symmetry and thereby enable conventional electric-dipole shift and injection currents, including spin photocurrents. In contrast, here the crystal may remain inversion symmetric even if magnetic moments are considered: the current instead becomes symmetry allowed because one of the two optical vertices is axial. Thus the mechanism does not rely on structural inversion breaking, interfacial polarity, or an applied dc magnetic field.

In summary, we have developed a unified microscopic framework for charge and spin photocurrents generated by interference between the electric- and magnetic-dipole couplings of light. The EMDI susceptibility is even under spatial inversion and therefore survives in centrosymmetric crystals, while time reversal separates shift-like and injection-like responses into complementary charge and spin sectors. In nonmagnetic centrosymmetric materials like bilayer MoS$_2$, only $C$-shift and $L$-injection charge channels exist. In constast, in magnetic materials like MnBi$_2$Te$_4$, all four channels, $C$-shift, $C$-injection, $L$-shift, and $L$-injection, survive. These results extend nonlinear photocurrent theory from charge transport to a unified charge-spin response and identify centrosymmetric quantum materials as a platform for optical spin-current generation without structural inversion breaking.

\bibliography{ref}

@article{alex2011tip,
  title={Tip-enhanced photovoltaic effects in bismuth ferrite},
  author={Alexe, M. and Hesse, D.},
  journal={Nat. Commun.},
  volume={2},
  pages={256},
  year={2011},
  doi = {10.1038/ncomms1261}
}

@article{bhatnagar2013role,
  title={Role of domain walls in the abnormal photovoltaic effect in {BiFeO3}},
  author={Bhatnagar, A. and Roy Chaudhuri, A. and Kim, Y. H. and Hesse, D. and Alexe, M.},
  journal={Nat. Commun.},
  volume={4},
  pages={2835},
  year={2013},
  doi = {10.1038/ncomms3835}
}

@article{spanier2016power,
  title={Power conversion efficiency exceeding the Shockley--Queisser limit in a ferroelectric insulator},
  author={Spanier, J. E. and Fridkin, V. M. and Rappe, A. M. and others},
  journal={Nat. Photonics},
  volume={10},
  pages={611--616},
  year={2016},
  doi = {10.1038/nphoton.2016.143}
}

@article{osterhoudt2019colossal,
  title={Colossal mid-infrared bulk photovoltaic effect in a type-I Weyl semimetal},
  author={Osterhoudt, G. B. and Diebel, L. K. and Gray, M. J. and others},
  journal={Nat. Mater.},
  volume={18},
  pages={471--475},
  year={2019},
  doi = {10.1038/s41563-019-0297-4}
}

@article{ma2019nonlinear,
  title={Nonlinear photoresponse of type-II Weyl semimetals},
  author={Ma, J. and Gu, Q. and Liu, Y. and others},
  journal={Nat. Mater.},
  volume={18},
  pages={476--481},
  year={2019},
  doi = {10.1038/s41563-019-0296-5}
}

@article{fridkin2001bulk,
  title={Bulk photovoltaic effect in noncentrosymmetric crystals},
  author={Fridkin, VM},
  journal={Crystallogr. Rep.},
  volume={46},
  number={4},
  pages={654--658},
  year={2001},
  doi = {10.1134/1.1387133},
  publisher={Springer}
}

@article{glass1995high,
  title   = {High-voltage bulk photovoltaic effect and the photorefractive process in {LiNbO3}},
  author  = {Glass, A. M. and von der Linde, D. and Negran, T. J.},
  journal = {Appl. Phys. Lett.},
  volume  = {25},
  pages   = {233--235},
  year    = {1974},
  doi     = {10.1063/1.1655453}
}

@article{von1981theory,
  title   = {Theory of the bulk photovoltaic effect in pure crystals},
  author  = {von Baltz, Ralph and Kraut, Wolfgang},
  journal = {Phys. Rev. B},
  volume  = {23},
  pages   = {5590},
  year    = {1981},
  doi     = {10.1103/PhysRevB.23.5590}
}

@article{sipe2000second,
  title={Second-order optical response in semiconductors},
  author={Sipe, JE and Shkrebtii, AI},
  journal={Phys. Rev. B},
  volume={61},
  number={8},
  pages={5337},
  year={2000},
  publisher={APS},
  doi = {10.1103/PhysRevB.61.5337}
}

@article{matsyshyn2021rabi,
  title={Rabi regime of current rectification in solids},
  author={Matsyshyn, Oles and Piazza, Francesco and Moessner, Roderich and Sodemann, Inti},
  journal={Phys. Rev. Lett.},
  volume={127},
  number={12},
  pages={126604},
  year={2021},
  publisher={APS},
  doi = {10.1103/PhysRevLett.127.126604}
}

@article{young2012first,
  title={First principles calculation of the shift current photovoltaic effect in ferroelectrics},
  author={Young, Steve M and Rappe, Andrew M},
  journal={Phys. Rev. Lett.},
  volume={109},
  number={11},
  pages={116601},
  year={2012},
  publisher={APS},
  doi = {10.1103/PhysRevLett.109.116601}
}

@article{sipe1993nonlinear,
  title   = {Nonlinear optical response of semiconductors in the independent-particle approximation},
  author  = {Sipe, J. E. and Ghahramani, E.},
  journal = {Phys. Rev. B},
  volume  = {48},
  pages   = {11705--11722},
  year    = {1993},
  doi     = {10.1103/PhysRevB.48.11705}
}

@article{bhalla2023quantum,
  title   = {Quantum kinetic theory of nonlinear optical currents: Finite Fermi surface and Fermi sea contributions},
  author  = {Bhalla, Pankaj and Das, Kamal and Agarwal, Amit and Culcer, Dimitrie},
  journal = {Phys. Rev. B},
  volume  = {107},
  pages   = {165131},
  year    = {2023},
  doi     = {10.1103/PhysRevB.107.165131}
}

@article{nastos2010optical,
  title={Optical rectification and current injection in unbiased semiconductors},
  author={Nastos, F and Sipe, JE},
  journal={Phys. Rev. B},
  volume={82},
  number={23},
  pages={235204},
  year={2010},
  publisher={APS},
  doi = {10.1103/PhysRevB.82.235204}
}

@article{dejuan2017quantized,
  title={Quantized circular photogalvanic effect in Weyl semimetals},
  author={de Juan, F. and Grushin, A. G. and Morimoto, T. and Moore, J. E.},
  journal={Nat. Commun.},
  volume={8},
  pages={15995},
  year={2017},
  doi = {10.1038/ncomms15995}
}

@article{ahn2020low,
  title={Low-frequency divergence and quantum geometry of the bulk photovoltaic effect in topological semimetals},
  author={Ahn, J. and Guo, G.-Y. and Nagaosa, N.},
  journal={Phys. Rev. X},
  volume={10},
  pages={041041},
  year={2020},
  doi = {10.1103/PhysRevX.10.041041}
}

@article{ahn2022riemannian,
  title={Riemannian geometry of resonant optical responses},
  author={Ahn, J. and Guo, G.-Y. and Nagaosa, N. and Vishwanath, A.},
  journal={Nat. Phys.},
  volume={18},
  pages={290--295},
  year={2022},
  doi = {10.1038/s41567-021-01465-z}
}

@article{ma2021topology,
  title={Topology and geometry under the nonlinear electromagnetic spotlight},
  author={Ma, Q. and Grushin, A. G. and Burch, K. S.},
  journal={Nat. Mater.},
  volume={20},
  pages={1601--1614},
  year={2021},
  doi = {10.1038/s41563-021-00992-7}
}

@article{ma2023photocurrent,
  title={Photocurrent as a multiphysics diagnostic of quantum materials},
  author={Ma, Q. and Krishna Kumar, R. and Xu, S.-Y. and Koppens, F. H. L. and Song, J. C. W.},
  journal={Nat. Rev. Phys.},
  volume={5},
  pages={170--184},
  year={2023},
  doi = {10.1038/s42254-022-00551-2}
}

@article{zhang2019symmetry,
  title={Symmetry-breaking-induced nonlinear optics at a microcavity surface},
  author={Zhang, Xueyue and Cao, Qi-Tao and Wang, Zhuo and Liu, Yu-xi and Qiu, Cheng-Wei and Yang, Lan and Gong, Qihuang and Xiao, Yun-Feng},
  journal={Nat. Photonics},
  volume={13},
  number={1},
  pages={21--24},
  year={2019},
  publisher={Nature Publishing Group UK, London},
  doi = {10.1038/s41566-018-0297-y}
}

@article{cook2017design,
  title={Design principles for shift current photovoltaics},
  author={Cook, Ashley M and M. Fregoso, Benjamin and De Juan, Fernando and Coh, Sinisa and Moore, Joel E},
  journal={Nat. Commun.},
  volume={8},
  number={1},
  pages={14176},
  year={2017},
  publisher={Nature Publishing Group UK, London},
  doi = {10.1038/ncomms14176}
}

@article{morimoto2016topological,
  title={Topological nature of nonlinear optical effects in solids},
  author={Morimoto, Takahiro and Nagaosa, Naoto},
  journal={Sci. Adv.},
  volume={2},
  number={5},
  pages={e1501524},
  year={2016},
  publisher={American Association for the Advancement of Science},
  doi = {10.1126/sciadv.1501524}
}

@article{tan2016shift,
  title={Shift current bulk photovoltaic effect in polar materials—hybrid and oxide perovskites and beyond},
  author={Tan, Liang Z and Zheng, Fan and Young, Steve M and Wang, Fenggong and Liu, Shi and Rappe, Andrew M},
  journal={npj Comput. Mater.},
  volume={2},
  number={1},
  pages={16026},
  year={2016},
  publisher={Nature Publishing Group},
  doi = {10.1038/npjcompumats.2016.26}
}

@article{taghinejad2020transient,
  title={Transient second-order nonlinear media: Breaking the spatial symmetry in the time domain via hot-electron transfer},
  author={Taghinejad, Mohammad and Xu, Zihao and Lee, Kyu-Tae and Lian, Tianquan and Cai, Wenshan},
  journal={Phys. Rev. Lett.},
  volume={124},
  number={1},
  pages={013901},
  year={2020},
  publisher={APS},
  doi = {10.1103/PhysRevLett.124.013901}
}

@article{morimoto2018current,
  title={Current-voltage characteristic and shot noise of shift current photovoltaics},
  author={Morimoto, Takahiro and Nakamura, Masao and Kawasaki, Masashi and Nagaosa, Naoto},
  journal={Phys. Rev. Lett.},
  volume={121},
  number={26},
  pages={267401},
  year={2018},
  publisher={APS},
  doi = {10.1103/PhysRevLett.121.267401}
}

@article{lihm2022comprehensive,
  title={Comprehensive theory of second-order spin photocurrents},
  author={Lihm, Jae-Mo and Park, Cheol-Hwan},
  journal={Phys. Rev. B},
  volume={105},
  number={4},
  pages={045201},
  year={2022},
  publisher={APS},
  doi = {10.1103/PhysRevB.105.045201}
}

@article{xu2021pure,
  title={Pure spin photocurrent in non-centrosymmetric crystals: bulk spin photovoltaic effect},
  author={Xu, Haowei and Wang, Hua and Zhou, Jian and Li, Ju},
  journal={Nat. Commun.},
  volume={12},
  number={1},
  pages={4330},
  year={2021},
  publisher={Nature Publishing Group UK, London},
  doi = {10.1038/s41467-021-24541-7}
}

\end{document}